\documentstyle[12pt,epsf]{article}
\def\QF{Quantum Fireball}
\def\issue(#1,#2,#3){\space{\bf{#1}}\space(#2)\space#3}
\def\NIM(#1,#2,#3){ Nuclear Instruments and Methods\issue(#1,#2,#3)}
\begin{document}
\begin{table}[t!]
\begin{flushright}
\begin{tabular}{l}
UMS/HEP/99--025 \\
FERMILAB-Conf-99-212 \\
\end{tabular}
\end{flushright}
\end{table}
\centerline{\Large Working with Arrays of Inexpensive EIDE Disk Drives}
\centerline{\Large (Including an Appendix with a December 1999 Update)}
\bigskip
\bigskip
\bigskip
\centerline{\large David Sanders, Chris Riley, Lucien Cremaldi, and Don Summers}
\vskip 1pt
\centerline{University of Mississippi--Oxford}
\vskip 8pt
\centerline{\large Don Petravick}
\vskip 1pt
\centerline{Fermilab}
\bigskip
\bigskip
\bigskip
\bigskip
\leftline{Abstract:}
\vskip 5pt
\parbox{5.5in}{
{\small \parindent=0pt
In today's marketplace, the cost per Terabyte of disks with EIDE interfaces is
about a third that of disks with SCSI.  Hence, three times as many particle
physics events could be put online with EIDE.  The modern EIDE interface
includes many of the performance features that appeared earlier in SCSI.  EIDE
bus speeds approach 33 Megabytes/s and need only be shared between two disks
rather than seven disks.  The internal I/O rate of very fast (and expensive)
SCSI disks is only 50 per cent greater than EIDE disks.  Hence, two EIDE disks
whose combined cost is much less than one very fast SCSI disk can actually
give more data throughput due to the advantage of multiple spindles and head
actuators.  We explore the use of 12 and 16 Gigabyte EIDE disks with
motherboard and PCI bus card interfaces on a number of operating systems and
CPUs.  These include Red Hat Linux and Windows 95/98 on a Pentium, MacOS and
Apple's Rhapsody/NeXT/UNIX on a PowerPC, and Sun Solaris on a UltraSparc 10
workstation.}}
\vfill
\centerline{Computing in High Energy Physics Conference (CHEP\,'98)}
\vskip 1pt
\centerline{August 31 -- September 4, 1998}
\vskip 1pt
\centerline{Hotel Inter--Continental}
\vskip 1pt
\centerline{Chicago, Illinois, USA}
\bigskip
\centerline{Contact: \ sanders@relativity.phy.olemiss.edu}
\bigskip
\centerline{This work was supported by the U.S.\ Department of Energy under}
\centerline{grant DE-FG02-91ER40622 and contract DE-AC02-76CH03000.} 
\eject

\centerline{\bf \Large Introduction}
\vskip 8pt
In today's marketplace, the cost per Terabyte of disks with EIDE (Enhanced 
Integrated Drive Electronics) interfaces is about a third that of disks with
SCSI (Small Computer System Interface).  Hence, three times as many particle
physics events could be put online with EIDE.  The modern EIDE interface
includes many of the performance features that appeared earlier in SCSI.  EIDE
bus speeds approach 33 Megabytes/s and need only be shared between two EIDE 
disks
rather than seven SCSI disks.  The internal I/O rate of very fast 
(and expensive)
SCSI disks is only 50 percent greater than EIDE disks.  Direct Memory Access
(DMA), scatter/gather data transfers without intervention of the Central
Processor Unit (CPU), elevator seeks, and command queuing are now available
for EIDE, as well as support for disks larger than 8.4 Gigabytes.  PCI
(Peripheral Control Interface) cards allow the addition of even more EIDE
interfaces, in addition to those already on the
motherboard.

\vskip 20pt
\centerline{\Large Motivation}
\vskip 8pt
There are a number of High Energy Physics Experiments that have produced 
Terabytes of data [1].  A few examples as of 12/95 are:

\begin{table}[h!]
\begin{center}
\renewcommand{\arraystretch}{1.15}
\begin{tabular}{|l|c|} \hline 
{\bf Experiment} & {\bf Data set (Terabytes)} \\ \hline \hline 
FNAL-E791  & 50  \\ \hline
FNAL-D0    & 40  \\ \hline
FNAL-CDF   & 10  \\ \hline
HERA-ZEUS  & 5   \\ \hline
CESR-CLEO  & 5   \\ \hline
LEP-Delphi & \~5 \\ \hline
LEP-L3     & 3.4 \\ \hline
HERA-H1    & 2.5 \\ \hline
LEP-Aleph  & 1.7 \\ \hline
LEP-OPAL   & 1.5 \\ \hline 
\end{tabular}
\end{center}
\end{table}

The efficiency of data analysis is greatly enhanced by using disk based files
of filtered Data Summary Tapes (DSTs) rather than continually loading files
from tapes. However, the high cost of disks have hindered more widespread use.
Low cost EIDE disks are improving this situation.
\vfill
\eject

\centerline{\Large Big Disks}
\vspace*{-15pt}
\begin{table}[h!]
\begin{flushleft}
\renewcommand{\arraystretch}{1.15}
\begin{tabular}{|l|c|c|c|c|} \hline 
{\bf EIDE} Disk Model & Bigfoot & Deskstar & Deskstar & Diamond \\ \hline
Manufacturer          & Quantum & IBM      & IBM      & Maxtor \\ \hline \hline
Capacity (Gigabytes)  & 12      & 16.8     & 14.4     & 17.2   \\ \hline
Max. Internal I/O (Mbits/s) & 142  & 162  & 174 & \\ \hline
Avg. Seek Time (ms)     & 12.0 & 9.5  & 9.5 & 9.0 \\ \hline
RPM                     & 4000 & 5400 & 7200 & 5400 \\ \hline
Unit Street Cost        & \$241 & \$407 & \$407 & \$410 \\ \hline
Cost \$/Terabyte  & \$20000 & \$24000 & \$28000 & \$24000 \\ \hline
\end{tabular}
\end{flushleft}
\end{table}
\vspace*{-30pt}
\begin{table}[h!]
\begin{flushleft}
\renewcommand{\arraystretch}{1.15}
\begin{tabular}{|l|c|c|} \hline 
{\bf SCSI} Disk Model & Cheetah & Ultrastar \\ \hline
Manufacturer  & Seagate  & IBM \\ \hline \hline
Capacity (Gigabytes) & 18.2 & 18 \\ \hline
Max. Internal I/O (Mbits/s) & 231 & 180 \\ \hline
Avg. Seek Time (ms)  & 5.7  & 6.5   \\ \hline
RPM        & 10000   & 7200 \\ \hline
Unit Street Cost & \$1242 & \$1000 \\ \hline
Cost \$/Terabyte & \$68000  & \$55000 \\ \hline
\end{tabular}
\end{flushleft}
\end{table}

\vskip 5pt
\centerline{\Large Tests Performed}

\vskip 8pt
For this paper we tested two of the large capacity EIDE disks with six
different operating systems and a PCI EIDE disk controller card.  The six
operating systems are Mac OS 8.1, Apple Rhapsody DR2, Sun Solaris 2.6, Windows
95b, Windows 98, and RedHat LINUX 5.1 (kernel 2.0.34).  The two disk drives
and the disk controller card are described below:

\begin{itemize}
\begin{raggedright}
\item Quantum Bigfoot$^{TM}$ TX [2]  12 GB, 4000 RPM, 
142 Mbits/sec 
Maximum internal data rate, 12 ms average seek time.

\item The IBM Deskstar$^{TM}$ 16GP [3].  16.8 GB, 5400 RPM, 
162 Mbits/sec 
Maximum internal data rate, 9.5 ms average seek time.

\item Promise Technologies Ultra 33$^{TM}$ PCI EIDE controller 
card [4].  
Supports 4
drives, Ultra ATA/EIDE/Fast ATA-2.  Cost: \$50.
\end{raggedright}
\end{itemize}

Both the Quantum Bigfoot$^{TM}$ TX 12 GB and the IBM Deskstar$^{TM}$ 16GP 16 GB 
disks were successfully tested with the following systems:
\vfill \eject
\null
\vspace*{-50pt}
\begin{table}[h!]
\begin{center}
\renewcommand{\arraystretch}{1.10}
\begin{tabular}{|l|l|} \hline 
{\bf System} & {\bf Notes} \\ \hline \hline
Mac OS 8.1 on a Macintosh G3 [5] rev. 2 & With HFS+ and both Master/Slave. \\ 
motherboard & \\ \hline
Windows 95b on a Dell Dimension XPS 350 & Ok, depending on the 
BIOS\footnotemark[1]. \\
computer [6]  with PhoenixBIOS              & Use FAT 32. \\ \hline
Windows 98 on a Dell Dimension XPS 350  & Ok, depending on the 
                        BIOS\footnotemark[1]. \\
computer [6] with PhoenixBIOS              & Use FAT 32. \\ \hline
RedHat LINUX 5.1 (kernel 2.0.34) on a Dell & Ok, depending on the 
                                              BIOS\footnotemark[1]. \\ 
Dimension XPS 350 computer [6] with  & \\ 
PhoenixBIOS & \\ \hline
Promise Technologies Ultra 33 on a Dell & Ok with Windows 95b and Windows \\
Dimension XPS 350 computer [6] with & 98. However, a patch\footnotemark[2]
was needed for \\ 
PhoenixBIOS  & Red Hat LINUX. \\ \hline
\end{tabular}
\end{center}
\end{table}

\vspace*{-2pt}
\centerline{\Large Ten Terabyte EIDE Disk Architecture}

\vskip 5pt
The recipe for a simple 10 Terabyte EIDE Disk Architecture is as follows:
{\itemsep=-2pt \parsep=-2pt
\begin{itemize}
\begin{raggedright}

\item  Attach eight 16GB EIDE disks to each of 75 CPUs with the help of 
   Promise PCI controller cards.
\vspace*{-5pt}
\item Since EIDE cables have a maximum length of 18", it is easier to run 
      extra DC power cables into a computer tower than to run EIDE cables out.
\vspace*{-5pt}
\item  Load data on these disk arrays.
\vspace*{-5pt}
\item Plan to usually run analysis jobs on the same machine as the data.
\vspace*{-5pt}
\item Use fast Ethernet switches to allow for remote jobs at a modest level.
\end{raggedright}
\end{itemize}}
\vskip 9pt
\centerline{\Large Future}
\vskip 5pt
Future plans may include testing the drives with Apple Rhapsody, Sun Solaris,
and newer releases of Red Hat LINUX.  (The 8
GB limit seen so far on Rhapsody DR2 and Solaris 2.6
may be fixed in later releases.)
Also new
technologies that are worth investigating include both ``Lazy RAID'' and
Firewire$^{TM}$.
\vskip 5pt
\footnotesep=5pt
\footnotetext[1]{
See ``Getting beyond the ATA 8.4 GB limit'' \hfil \break
http://www.storage.ibm.com/hardsoft/diskdrdl/library/8.4gb.htm \hfil \break
http://www.storage.ibm.com/techsup/hddtech/welcome.htm         \hfil \break
http://www.storage.ibm.com/hardsoft/diskdrdl/prod/deskstar.htm \hfil \break
http://www.storage.ibm.com/hardsoft/diskdrdl/prod/ultrastar.htm \hfil \break
and ``8.4 GB Barrier'' \hfil \break
http://www.quantum.com/src/whitepapers/8.4barrier.html \hfil \break
http://www.quantum.com/products/hdd/bigfoot\_tx/ \hfil \break
and ``IDE Hard Drive Capacity Barriers'' \hfil \break
http://www.maxtor.com/technology/whitepapers/capbar0.html}
\footnotetext[2]{ 
Patch available from http://pobox.com/~brion/linux/promise34.gz, \hfil \break
but support is included in kernel 2.0.35}

\eject
\centerline{\Large Lazy RAID}
\vskip 5pt
Lazy RAID (Redundant Array of Inexpensive Disks) is an idea for using disk 
arrays that offers protection for disks in the event of catastrophic
failure of one disk in the array.  This system uses a number of data disks
(say 7) plus one parity disk. Therefore, if one disk dies the parity disk
would allow the recovery of data from the dead disk.  One could use the RAW
DEVICE interface to calculate parity with the CPU.  If a disk fails then the
operator would swap out the dead EIDE drive and reconstitute the dead disk
drive onto the replacement drive using the parity disk and the remaining data
disks. This system is well suited for use as scratch disks where a filtered
DST is placed on disk once and read and analyzed many times.  Using this
scheme the one parity disk is updated
only when a file is written to (or erased from) a disk.
\vskip 8pt
\centerline{\Large Firewire}
\vskip 5pt
Firewire IEEE 1394 Specifications [7]:
{\itemsep=0pt \parsep=0pt
\begin{itemize}
\begin{raggedright}
\item Up to 25 or 50 Megabyte/s.
\item Up to 63 devices per interface.
\item Uses two twisted pair data lines.
\item  ``Fairness'' bus arbitration.
\item Supported by MacOS and Windows 98.
\end{raggedright}
\end{itemize}}
A printed circuit board and DSP driver software would have to be developed
using the TI chip set .  Shown below is a Firewire to EIDE Disk Block Diagram
that might allow one Terabyte Per PCI Slot:
 
\begin{figure}[h!]
\centerline{\epsfxsize 5.5 truein \epsfbox{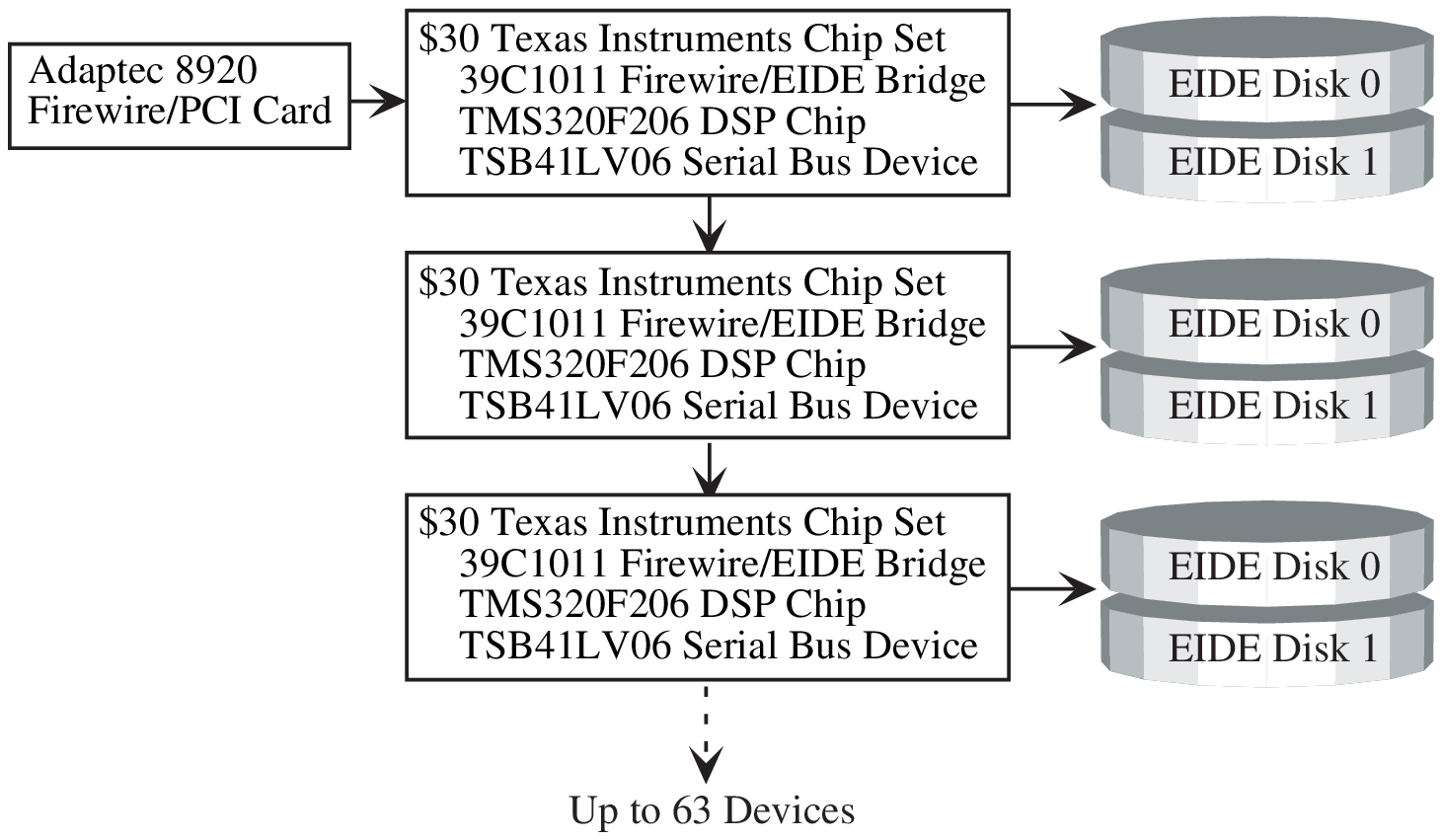}}
\label{Firewire}
\end{figure}

\vfill \eject
\centerline{\Large Conclusion}
\vskip 5pt
EIDE disk arrays are an inexpensive way to add large amounts of disk space to
both single Workstations (and PCs) and multiprocessor computing farms.  They
provide an additional layer to the data storage ``cake''.

\vskip 5pt
\begin{figure}[h!]
\centerline{\epsfxsize 5.5 truein \epsfbox{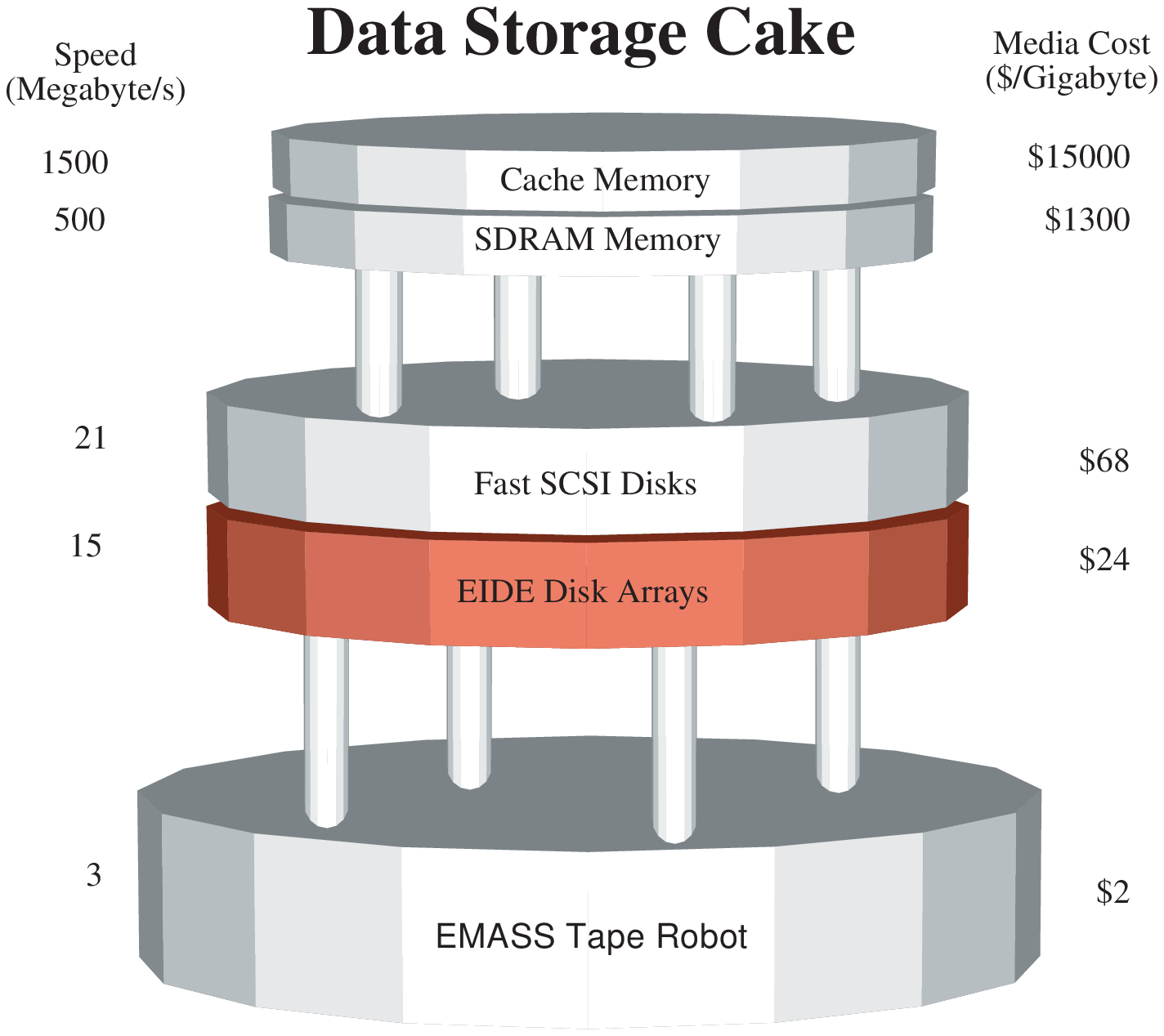}}
\label{Cake}
\end{figure}

\vskip 20pt 
\footnoterule
\vspace*{-10pt}
\begin{enumerate}
\begin{raggedright}
\parsep 0pt
\itemsep 0pt
{\footnotesize
\item[{[1]}] S.~Bracker, K.~Gounder, K.~Hendrix, and D.~Summers, {\em A Simple
Multiprocessor Management System for Event-Parallel Computing,} IEEE NS-43
(1996) 2457.
\item[{[2]}] http://www.quantum.com/products/hdd/bigfoot\_tx/
\item[{[3]}] http://www.storage.ibm.com/hardsoft/diskdrdl/desk/1614data.htm
\item[{[4]}] http://www.promise.com/html/sales/Ultra33.html
\item[{[5]}] http://www.apple.com/powermac/g3/
\item[{[6]}] http://www.dell.com/products/dim/xpsr/specs/index.htm
\item[{[7]}] http://www.skipstone.com/info.html
\item[{[8]}] http://www.ti.com/sc/docs/news/1998/98029.htm; \hfil \break
http://www.ti.com/sc/docs/dsps/details/43/flash.htm; \hfil \break
http://www.ti.com/sc/docs/msp/1394/41lv0x.htm;             \hfil \break
http://www.ti.com/sc/docs/storage/products/cont.htm
}
\end{raggedright}
\end{enumerate}
\vfill
\eject
\centerline{\Large EIDE/ATA Disk Drive Update --- December 1999}
\vskip 10pt

Costs of disk drives with EIDE/ATA interfaces [1] have fallen since the Chicago
CHEP\,'98 conference. Drive capacity and I/O rates have risen. EIDE/ATA disks
beyond the old 8 Gigabyte limit now work with more platforms and cards. EIDE
disks remain more than twice as cost effective as SCSI disks and now equal the
internal I/O speeds of many SCSI disks.  We have run RAID 5 on EIDE disks
under Linux which both stripes data across disks for speed and provides parity
bits for data recovery. Tape backup may no longer be 
required to recover from one
disk failure in a set. Arrays of EIDE disks with Linux PCs serving as disk
controllers are attractive and may 
provide many Terabytes of economical rotating online storage. 
The cost of a quarter
Petabyte EIDE disk farm is approaching the cost of a StorageTek
PowderHorn silo [2] with 5000 50 Gigabyte RedWood tapes. Finally, we include
more information on IEEE--1394 FireWire which may allow Terabyte arrays of
EIDE/ATA disks to be directly connected to one or more computers at up to 50
Megabytes/second per interface.

\vfill
\begin{table}[h!]
\begin{center}
\renewcommand{\arraystretch}{1.08}
\tabcolsep=1.0mm
\begin{tabular}{lcccccc} \hline \hline
Table 1.       &  Disk     &  Maximum &     &      & Unit   &        \\
               &  Capacity &  Internal &  Avg. &    & Street & Dollars \\
{\bf EIDE/ATA} &      in   &       I/O & Seek  &     &  Cost  &  per    \\
Disk Drive Model & Gigabytes & Mbits/s &  ms & RPM  & [3]  & Terabyte \\ \hline
Maxtor DiamondMax 40 [4] & 40.9  &  295  & 9.0 & 5400 & \$243 &  \$6000 \\
IBM Deskstar 37GP [5] &    37.5  &  248  & 9.0 & 5400 & \$302 &   \$8100 \\
IBM Deskstar 34GXP [5] &   34.2  &  284  & 9.0 & 7200 & \$311 &  \$9100 \\
Seagate Barracuda ATA [6] & 28.5 &  323  & 8.6 & 7200 & \$211 &   \$7400   \\
Western Digital Expert [7] & 27.2 & 284  & 9.0 & 7200 & \$236 &   \$8700 \\
Western Digital Caviar [8] & 30.7 & 271  & 9.5 & 5400 &  *    &   *     \\
Quantum Fireball {\it lct10} [9] & 30.0 & 297 & 8.9 & 5400 & * & * \\ 
Maxtor DiamondMax Plus [10] & 40.9  &  345  & 9.0 & 7200 &  * &  *  \\
\hline 
& & & & & & \\
& & & & & & \\
\hline 
Table 2.    &  Disk     &  Maximum &     &      & Unit   &        \\
            &  Capacity &  Internal &  Avg. &     & Street & Dollars \\
{\bf SCSI}      &      in   &       I/O & Seek  &     &  Cost  &  per    \\
Disk Drive Model & Gigabytes & Mbits/s &  ms & RPM & [2]   & Terabyte \\ \hline
Seagate Barracuda [11]   & 50.1 & 264 & 7.4 &    7200 &  \$806 &  \$16\,100 \\ 
Quantum Atlas 10K [12]  & 36.4 & 315 & 5.0 & 10000 &    \$945 &  \$26\,000 \\
IBM Ultrastar 36XP [13] & 36.4 & 231 & 7.5 &    7200 &  \$756 &  \$20\,800 \\
IBM Ultrastar 72ZX [14] & 73.4 & 473 & 5.3 & 10000 &  * &  * \\
Seagate Cheetah 73 [15]   & 73.4 & 427 & 5.6 & 10000 &   * &  * \\
\hline \hline
\multicolumn{2}{l}{\footnotesize * Announced, but not yet shipping.} 
& & & & & \\
\end{tabular}
\end{center}
\end{table}
\eject
We now have our 12 GB Quantum Bigfoot TX [16] and 16.8 GB IBM Deskstar 16GP [17]
EIDE drives running in our Sun Ultra 10 workstation.  We put in a newer
motherboard (Sun Part No.\ 375-0009-09,  Date Code: 9843  \ DARWIN M/B) and
upgraded the operating system from Solaris 2.6 to Solaris 7. We did not test
the two changes individually, but one or both put us over the old 8 Gigabyte
limit.

Promise Technology now sells their Ultra66 [18] EIDE to PCI controller card.
The Ultra66 provides up to 66 MB/s on each of two channels with two drives per
channel. The card costs \$40. We use Promise's previous 33 MB/s EIDE to
PCI card daily in our Linux PC server (mail, backup...)
with IBM Deskstar disks [19].

ProMAX [20] sells a TurboMAX/ATA 33 Host Adapter for Macintosh PCI
buses. It allows adding four EIDE drives to an Apple Macintosh. Disks can be
striped in pairs.  FirmTek [21] is working on an Apple Macintosh software
driver for the Promise Ultra66 EIDE to PCI card.

\vskip 15pt
\leftline{\bf Terabytes of Linux RAID 5 Disks}

RAID [22] stands for Redundant Array of Inexpensive Disks.  Many industry
offering meet all of the qualifications except the inexpensive part, severely
limiting the size of an array for a given budget. This may change. RAID on
EIDE disks under Linux software which both stripes data across disks for speed
and provides parity bits for data recovery (RAID 5) is now available [23].
With redundant disk arrays, tape backup is not needed to recover from 
the failure of one disk in a set.
This removes a major obstacle to building large arrays of EIDE disks.  A RAID
5 set of eight 41 Gigabyte disks fits in a full tower case of a PC running
Linux.  This provides over a quarter of a Terabyte per box.  The boxes would be
connected using 100 Megabit/second Fast Ethernet PCI cards in each box plus
Ethernet switches [24, 25]. This looks to be very doable.

We have done a quick test of the Linux RAID 5 software using two 25 Gigabyte
IBM Deskstar 25GP [19] EIDE disks.  The host was a Pentium II with Red Hat
Linux 6.0 [26] and a Promise Technology Ultra 33 EIDE/PCI card.  The test ran
as expected. Naturally, half the disk space is devoted to parity with only two
disks. For a real RAID 5 system, eight or more disks would be a more efficient
use of space. The fraction of disk space devoted to parity equals the inverse
of the number of disks in a set.

\setcounter{table}{2}
\begin{table}[ht!]
\caption{Power consumption of large EIDE disk drives.}
\begin{center}
\renewcommand{\arraystretch}{1.15}
\tabcolsep=1.5mm
\begin{tabular}{lcccccc} \hline \hline
                    &          &     &     &      & Read/  &        \\
                    & & \multicolumn{2}{c}{Startup}& & Write/ &        \\
{\bf EIDE/ATA}      &           &  5V & 12V &Seek  &  Idle  & Standby \\
Disk Drive Model    &  GB       & Amps& Amps&Watts &  Watts & Watts \\  \hline
Maxtor DiamondMax 40 [4] & 40.9 & 0.3 & 2.1 & 11.0 &  5.2   & 1.3  \\
IBM Deskstar 37GP [5] &   37.5  & 1.0 & 2.0 &      &  4.9   &      \\
IBM Deskstar 34GXP [5] &  34.2  & 1.0 & 2.0 &      &  6.9   &      \\
Seagate Barracuda ATA [6] &28.5 & 0.7 & 2.4 & 9.7  &  7.0   & 0.8  \\
Western Digital Expert [7]&27.2 & 0.7 & 2.0 & 10.1 &  6.9   & 0.8  \\
Western Digital Caviar [8]&30.7 & 0.6 & 1.8 & 10.0 &  6.2   & 1.3  \\
\QF \, {\it lct10} [9]   & 30.0 & 0.7 & 1.7 &      &  5.0   &      \\ 
Maxtor DiamondMax Plus [10] & 40.9 &0.6& 2.5 & 12.9 &  6.7   & 1.4  \\
\hline \hline
\end{tabular}
\end{center}
\end{table}

On a mundane note, disk drives typically draw two amps at 12 volts for 15
seconds when starting. Thus eight drives can draw 16 amps at 12 volts.  This
can tax the ratings of an inexpensive commodity 300 watt PC power supply. Care
needs to be taken to choose a supply with a large portion of its wattage
devoted to 12 volts. The U.S.\ EPA Energy Star/Green PC Initiative has led to
the development of a Standby command for disks that might allow a staggered
startup of a disk array. The command ``/sbin/hdparm -S n" will spin down disks
under Linux. As array size grows, a second commodity 300 watt PC power supply
might be required per case.

\vfill
\eject
\leftline{\bf Comparison of Quarter Petabyte Disk and Tape Storage Systems}

In Table 4, we compare a quarter Petabyte EIDE disk farm to an automated
StorageTek PowderHorn tape silo [2] with eight RedWood tape drives and 5000 50
Gigabyte tapes. The disk farm estimate includes disks, parity disks, Linux and
RAID software, CPUs, motherboards, cases [27], power supplies, memory, Fast
Ethernet PCI cards, Promise Ultra66 cards, Ethernet switches [24, 25], and
racks. The Linux PC that runs each disk set costs about the same as a high end
SCSI-to-PCI controller card. 

To achieve a quarter Petabyte, 873 Linux PCs are required with eight 40.9 GB
disks each. One eighth of the disk space is devoted to parity for data
recovery from disk failure.  Care must be taken to write protect files and
disks to prevent accidental deletion. Physically, the PCs form a wall 4 high
by 2 deep by 110 wide (2.4 $\times$ 1.1 $\times$ 24 meters). Each Linux PC
consumes about 90 watts, equally divided between the disks and the
CPU/motherboard. A dozen 24\,000 BTU window air conditioners would suffice to
remove this 80 Kilowatt heat load. Much less heat is generated in standby
mode. The first level network consists of 288 \$75 fast ethernet switches
[25].  A single high end switch with 288 fast ethernet ports is used for the
network backbone [24]. 
The disks themselves can be used to transport data between sites.
A high rate experiment might generate a Terabyte of data a day which one
wished to move.
A Terabyte fits on 25 disks, which easily fit in a suitcase for shipping.

In summary, the disk farm  cost is not too much greater than
the tape silo cost and the performance of the disks is far better. One also
gets a Teraflop of computing power [28] as a free bonus; and the disk farm
encapsulates data with instructions in physical computing objects which can be
exploited to increase efficiency. Disk farm sizes can be scaled in size with
great flexibility! Its sometimes difficult for a university to buy a whole
tape silo [29].  Now everyone can have the benefit of online data.
\newpage

\begin{table}[ht!]
\caption{Comparison of a 250 TB EIDE disk array to a 250 TB 
StorageTek tape silo.}
\begin{center}
\renewcommand{\arraystretch}{1.15}
\tabcolsep=1.5mm
\begin{tabular}{lccccccc} \hline \hline  
         &         &       &      & No.~of   & I/O   & Time &   \\
         &         & Avg.  & Avg. & Disk or     & per  & to &   \\
Storage  & Vol.  & Power & Access  & Tape & Drive & Read & Purchase  \\
System   & m$^3$ & Kw & Time   & Drives & MB/s  & 250 TB & Price \\ \hline
EIDE RAID Disks & 100 & 80 & 0.009 s & 7000 & 20 & 34 minutes & \$2\,100\,000 \\
STK Tape Silo & 25  & 6 & 115 s & 8 & 11 & 32 days    & \$1\,300\,000  \\
\hline \hline
\end{tabular}
\end{center}
\end{table}

\leftline{\bf Terabyte Arrays of IEEE--1394 FireWire Disks}

   The amount of disk space one can connect to a CPU directly with 18" EIDE
cables is currently less than a Terabyte.  In some applications, one might
want to access more data with fast local disk and not suffer the overhead of
network software.  One may be able to use four IEEE--1394 FireWire buses [30] 
with a single CPU to attach up to 63 inexpensive EIDE disks per bus for a 
total of 10 Terabytes of local storage at 200 MB/s.
FireWire's peer--to--peer topology also adds significant new functionality by
allowing multiple computers to share multiple disks directly on a single bus.

Symbios Logic/LSI Logic has a \$13 IEEE--1394 to ATA/ATAPI controller chip 
[31]. The
SYM13FW500 integrates a 400 Mbits/s IEEE--1394 (FireWire) physical interface
(PHY) with an ATA/ATAPI interface, all on a low--power CMOS IC. Each
SYM13FW500 supports two ATA/ATAPI devices.  Wyle Electronics distributes 
the part.  
Recently, Oxford Semiconductor has also introduced
an IEEE--1394 to ATA/ATAPI controller chip, the OXFW900 [32]. 
Texas Instruments has decided not to market their prototype IEEE--1394 to
ATA/ATAPI chip.

EIDE disks with EIDE to IEEE 1394 FireWire interfaces are available from LaCie
[33] and VST Technologies[34].  VST Technologies has also shown
FireWire RAID arrays with mirroring and striping, but not parity [35].
All the disks are EIDE.

An interesting possibility might be to put eight EIDE drives in an inexpensive
PC case with a 300 watt power supply.  Then add four EIDE to FireWire interface
chips to a circuit board with the same form factor as a PC motherboard. The
color of the PC case can even be special ordered [27] for Apple Macintosh users.

Andreas Bombe, Sebastien Rougeaux, and Emanuel Pirker are in the process of
writing GNU Linux software drivers for IEEE 1394/FireWire devices [36].

Pieces appear to be converging.  It may, in the not too distant future, be
possible to directly connect Terabyte arrays of EIDE/ATA disks to one or more
computers at up to 50 Megabytes/second per FireWire interface.  Some computers
come with FireWire on the motherboard. FireWire interfaces can also be added
with cards such as OrangeMicro's HotLink FireWire PCI Board [37].

\vskip 10pt
\leftline{\bf REFERENCES}
\vspace*{-5pt}
\begin{enumerate}

\item[{[1]}] Friedhelm Schmidt, {\em The SCSI Bus and IDE Interface:
Protocols, Applications, and Programming,} Second edition, Addison--Wesley
(1997) ISBN 0-201-17514-2. EIDE stands for Enhanced Integrated 
Drive Electronics.

\item[{[2]}] 
http://www.storagetek.com/StorageTek/hardware/tape/9310/9310\_sp.html \hfil 
                                                                     \break
http://www.storagetek.com/StorageTek/hardware/tape/SD3/SD3\_sp.html
\item[{[3]}] http://www.shopper.com \hfil \break
             http://www.gateway.com/spotshop/ \ Validation Number KM120 \ \ 
                                                                       10\% off
\item[{[4]}] http://www.maxtor.com/diamondmax/40.html
\item[{[5]}] http://www.storage.ibm.com/hardsoft/diskdrdl/desk/37gp34gxpdata.htm
\item[{[6]}] {\small
http://www.seagate.com/cda/products/discsales/personal/family/0,1128,154,00.html
}
\item[{[7]}] http://www.westerndigital.com/company/releases/990426.html \hfil
                                                                        \break
             http://www.westerndigital.com/products/drives/specs/wd273bas.html 
\item[{[8]}] 
http://www.western-digital.com/products/drives/specs/wd307aas.html
\item[{[9]}] {\small
http://www.quantum.com/products/hdd/fireball\_lct10/fireball\_lct10\_specs.htm}
\item[{[10]}] http://www.maxtor.com/diamondmaxplus/40p.html
\item[{[11]}] 
http://www.seagate.com/products/discsales/discselect/A1a1.html \hfil \break
{\small
http://www.seagate.com/cda/products/discsales/enterprise/family/0,1130,43,00.html
}
\item[{[12]}] 
http://www.quantum.com/products/hdd/atlas\_10k/atlas\_10k\_specs.htm 
\hfil \break
http://www.quantum.com/products/hdd/atlas\_10kii/atlas\_10kii\_specs.htm
\item[{[13]}] http://www.storage.ibm.com/hardsoft/diskdrdl/ultra/36xpdata.htm
\item[{[14]}] http://www.storage.ibm.com/hardsoft/diskdrdl/ultra/72zxdata.htm
\item[{[15]}] 
http://www.seagate.com/products/discsales/discselect/A1a1.html \hfil \break
{\small
http://www.seagate.com/cda/products/discsales/enterprise/tech/0,1131,214,00.shtml
}
\item[{[16]}] {\small 
 http://www.quantum.com/products/archive/bigfoot\_tx/bigfoot\_tx\_overview.htm
}
\item[{[17]}] http://www.storage.ibm.com/hardsoft/diskdrdl/desk/1614data.htm
                                                                \hfil \break
              http://www.storage.ibm.com/hardsoft/diskdrdl/prod/14gx16gppr.htm
\item[{[18]}] http://www.promise.com/Products/idecards/u66.htm     \hfil \break
           http://www.promise.com/Products/idecards/u66compat.htm  \hfil \break
           http://www.promise.com/Latest/latedrivers.htm\#linuxu66 
\item[{[19]}] http://www.storage.ibm.com/hardsoft/diskdrdl/prod/ds25gp22.htm
\item[{[20]}] ProMAX, 16 Technology Dr., Irvine CA 92618  \ \ 800--977--6629 
              \hfil \break http://www.promax.com/
\item[{[21]}] http://www.firmtek.com \ Chi Kim Nguyen \ ckn@firmtek.com 
\item[{[22]}] David A.~Patterson, Garth Gibson, and Randy H.~Katz 
              (UC--Berkeley), {\em A Case for Redundant Arrays of Inexpensive 
              Disks (RAID),} Proceedings of the ACM Conference on Management 
              of Data (SIGMOD), Chicago, Illinois, USA (June 1988) 
              Pages 109--116, Sigmod Record {\bf 17} (1988) 109.
\item[{[23]}] Miguel de Icaza, Ingo Molnar, and Gadi Oxman, 
              {\em The LINUX RAID-1,4,5 Code,} 3rd Annual Linux Expo '97, 
         Research Triangle Park, North Carolina, USA (April 1997); \ \ 
         http://luthien.nuclecu.unam.mx/$\sim$miguel/raid    \hfil \break
         http://linas.org/linux/Software-RAID/Software-RAID.html   \hfil \break
         http://linas.org/linux/Software-RAID/Software-RAID-3.html
\item[{[24]}] We use the Lucent Cajun P550 Gigabit Switch with 23 Gigabits per 
              second of switching throughput capacity to connect our Fast 
              Ethernet computers. Up to six cards may be installed in this
              switch. One option is a card with 48
              full duplex 10/100Base-TX ports. \hfil \break
              http://public1.lucent.com/dns/products/p550.html
\item[{[25]}] One might quadruple the number of ports of a high end switch
like the Cajun P550 by adding a commodity switch to each of its ports.
Several Ethernet switches with 5 full duplex 10/100Base-TX ports cost under
\$100. All feature a store-and-forward packet switching architecture to help
reduce latency. \hfil \break
http://www.addtron.com/      \hfil    ADS--1005 \quad \$69 \break
http://www.dlink.com/products/switches/dss5plus/ \hfil DSS-5+ \quad \$90 \break 
http://www.hawkingtech.com/   \hfil PN505ES \quad  \$79 \break 
{\small 
http://www.linksys.com/scripts/features.asp?part=EZXS55W} \hfil EtherFast 
\quad \$90  \break
http://netgear.baynetworks.com/products/fs105ds.shtml \hfil  FS105  \quad
  \$83 \break
http://smc.com/smc/pages\_html/switch.html  \hfil  EZNET--5SW/6305TX \quad
 \$78 \break
\vspace*{-10pt}
\item[{[26]}] http://www.redhat.com 
\item[{[27]}] 
IW--Q600 ATX Full Tower Case, 11 bays, 300 watt power supply -- \$77 
                                                                  \hfil \break 
Q600 case dimensions: 600mm high $\times$ 200mm wide $\times$ 432mm deep. 
                                                                  \hfil \break 
IW--Q2000 ATX Full Tower Case, 11 bays, two 300 watt power supplies -- \$208 
                                                                  \hfil \break 
Q2000 case dimensions: 600mm high $\times$ 200mm wide $\times$ 476mm deep. 
                                                                  \hfil \break 
              ``Available in different color for OEM customers.'' \hfil \break 
              http://www.in-win.com/framecode/index.html  \hfil \break
              http://www.pricewatch.com
\item[{[28]}]
Advanced Micro Devices and Pentium III CPUs can perform four single precision
floating point adds or multiplies per clock cycle with their 3DNow!\ or
Streaming SIMD Extensions units, respectively. Both 3DNow!\ and SSE are
implementations of SIMD (Single Instruction, Multiple Data) processors. \hfil
\break
http://www.amd.com/products/cpg/athlon/index.html             \hfil \break
http://www.amd.com/products/cpg/k6iii/index.html              \hfil \break
http://www.amd.com/products/cpg/k623d/index.html              \hfil \break
http://www.intel.com/home/prodserv/pentiumiii/prodinfo.htm   
\item[{[29]}] 
For situations with more people than money, manually loaded tapes
provide the way to store and move data with the lowest initial investment. The
lowest media cost is given by 112 meter long 8mm tapes storing 5 Gigabytes
uncompressed on an Exabyte Eliant 820 at 1 MB/s. The tapes cost 53 cents per
Gigabyte at the Fermilab stockroom and the Eliant 820 tape drive costs \$1300.
Used Exabyte 8500 and 8505 tape drives are even cheaper on ebay.com. Using
pairs of drives, with one running and the other waiting on deck with a tape
ready to go, gives operators time to load tapes [38, 39].
                                                                   \hfil \break
http://www.exabyte.com/products/8mm/eliant/ \hfil \break
http://www-stock.fnal.gov/stock/            \hfil \break
http://www.exabyte.com/products/
\item[{[30]}] 
{\em IEEE Standard for a High Performance Serial Bus,}
ISBN 1--55937--583--3. \hfil \break
http://standards.ieee.org/catalog/bus.html\#1394-1995  \hfil \break
http://standards.ieee.org/reading/ieee/std\_public/description \hfil \break
/busarch/1394-1995\_desc.html
\item[{[31]}] 
SYM13FW500 ATA/ATAPI to 1394 Native Bridge Data Manual Version 1.02 \hfil \break
              ftp://ftp.symbios.com/pub/symchips/1394  \hfil \break 
              http://www.symbios.com/news/pr/80330ata.htm  
\item[{[32]}] http://www.oxsemi.com/products/products.html 
\item[{[33]}] http://www.lacie.com/scripts/harddrive/drive.cfm?which=30
\item[{[34]}] VST Technologies, 125 Nagog Park,  Acton, MA  01720 \ \ 
              978--635--8282 \hfil \break
              http://www.vsttech.com/vst/products.nsf/pl\_firewire \hfil \break
            http://www.macnn.com/thereview/reviews/vst/fwhd.shtml \hfil \break 
              http://www.elgato.com/products.html \quad FireWire Disk Control
              1.01
\item[{[35]}] http://www.vsttech.com/vst/press.nsf/default \ 08/31/99 \ 09/17/99 \hfil \break
              http://www.softraid.com
\item[{[36]}] http://eclipt.uni-klu.ac.at/ieee1394/  \hfil \break
              http://www.kt.opensrc.org/kt19990722\_28.html\#15
\item[{[37]}] http://www.orangemicro.com/firewire.html
\item[{[38]}] FNAL E769: C.~Stoughton and D.~Summers, 
{\em Using Multiple RISC 
CPUs in Parallel to Study Charm Quarks,} Comput.~Phys. {\bf 6} (1992)
371; C. Gay and S. Bracker, IEEE {\bf NS-34} (1987) 870;
S. Hansen et al., IEEE {\bf NS-34} (1987) 1003;
G.\,Alves et al., Phys.\,Rev.\,Lett. {\bf 69} (1992) 3147;
{\bf 77} (1996) 2388; {\bf 77} (1996) 2392;
Phys.\,Rev. {\bf D56} (1997) 6003.

\item[{[39]}] 
FNAL E791: S.~Bracker, K.~Gounder, K.~Hendrix, and D.~Summers, 
{\em A Simple
Multiprocessor Management System for Event-Parallel Computing,} IEEE 
{\bf NS-43} (1996) 2457; 
S. Amato et al., Nucl. Instrum. Meth. {\bf A324} (1993) 535; \\
E. M. Aitala et al., Phys.\,Rev.\,Lett. {\bf 76} (1996) 364;
{\bf 81} (1998) 44;
hep-ex/9809026; hep-ex/9809029; hep-ex 9912003.  

\end{enumerate}
\end{document}